\documentclass[%
 reprint,
showpacs,
 amsmath,amssymb,
 aps,
pre,
]{revtex4-1}

\usepackage{graphicx}
\usepackage{dcolumn}
\usepackage{bm}

\usepackage{color}
\usepackage[export]{adjustbox}

\begin{document}
\preprint{APS/123-QED}
\newcommand{\tbox}[1]{\mbox{\tiny #1}}

\title{Geometrical and spectral study of $\beta$-skeleton graphs}

\author{L. Alonso}
\email{lalonso@pks.mpg.de}
\affiliation{
Max-Planck-Institut f\"{u}r Physik komplexer Systeme, N\"{o}thnitzer Str. 38, 01187  Dresden, Germany
}

\author{J. A. M\'endez-Berm\'udez}
\email{jmendezb@ifuap.buap.mx}
\affiliation{
Instituto de F\'{i}sica, Benem\'erita Universidad Aut\'onoma de Puebla, Apartado Postal J-48, Puebla 72570, Mexico
}

\author{Ernesto Estrada}
\email{estrada66@unizar.es}
\affiliation{
Institute of Mathematics and Applications (IUMA), University of Zaragoza, Pedro Cerbuna 12, Zaragoza 50009, Spain  and ARAID Foundation, Government of Aragon, 50008 Zaragoza, Spain
}

\date{\today}

\begin{abstract}
We perform an extensive numerical analysis of $\beta$-skeleton graphs, a particular type of proximity graphs.
In a $\beta$-skeleton graph (BSG) two vertices are connected if a proximity rule, that depends of the parameter 
$\beta\in(0,\infty)$, is satisfied. Moreover, for $\beta>1$ there exist two different proximity rules, leading to
lune-based and circle-based BSGs. First, by computing the average degree of large ensembles of BSGs 
we detect differences, which increase with the increase of $\beta$, between lune-based and circle-based BSGs.
Then, within a random matrix theory (RMT) approach, we explore spectral and eigenvector properties 
of randomly weighted BSGs by the use of the nearest-neighbor energy-level spacing distribution 
and the entropic eigenvector localization length, respectively. The RMT analysis allows us to conclude 
that a localization transition occurs at $\beta=1$. 
\end{abstract}

\pacs{89.75.Hc, 64.60.-i, 05.45.Mt}
\keywords{Suggested keywords}
\maketitle


\twocolumngrid

\section{Introduction} 

The analysis of spatial networks plays a fundamental role for understanding complex systems embedded in geographical spaces, see~\cite{B11,B18}. Here we study a model which is a generalization of the so-called random neighborhood graphs~\cite{DGKJDR,GTT}, 
known as $\beta$-skeleton graphs, embedded in the unit square. In a $\beta$-skeleton graph (BSG) two vertices (points or nodes) are 
connected by an edge if and only if these vertices satisfy a particular geometrical requirement named as proximity 
rule. The proximity rule is encoded in the $\beta$ parameter which takes values in the interval $0<\beta<\infty$. 
With the proximity rules we will define below, a fully connected graph is obtained in the limit $\beta\rightarrow 0$,
while the network becomes a disconnected graph when $\beta\rightarrow\infty$.

In particular, BSGs are useful to study geometric complex systems where the connectivity between two items is interfered by the presence of a third one in between them. This is the case, for instance of granular materials~\cite{BOD12}, for representing urban street networks~\cite{OH14}, as well as for representing fractures in rocks~\cite{estrada}, among others.

This work is organized as follows.
In Sec.~II we introduce the proximity rules needed to construct the BSGs.
In fact, for $\beta>1$ there exist two different proximity rules, leading to lune-based and circle-based BSGs.
Indeed, our study focus on a detailed comparison between both.
Therefore, we study topological and spectral properties of BSGs by the use of 
the average degree, in Sec.~III, and nearest-neighbor energy-level spacing distribution 
and the entropic eigenvector localization length, in Sec.~IV.
Finally, we summarize in Sec.~V.

\section{Definitions of $\beta$-skeleton graphs}

For a given set of vertices $V =\{v_1, v_2,\dots, v_n\}$ on the plane, an Euclidean distance function $d$, and a parameter 
$0<\beta<\infty$, a graph $G_{\beta}(V)$, called BSG, is defined as follows~\cite{DGKJDR}:

{\it Two vertices $v_i, v_j \in V$ are connected with an edge iff no point from $V$\textbackslash$\{v_i, v_j\}$ 
belongs to the neighborhood ${\cal N}(v_i, v_j, \beta)$, where:}

\begin{figure}
\includegraphics[width = 5cm]{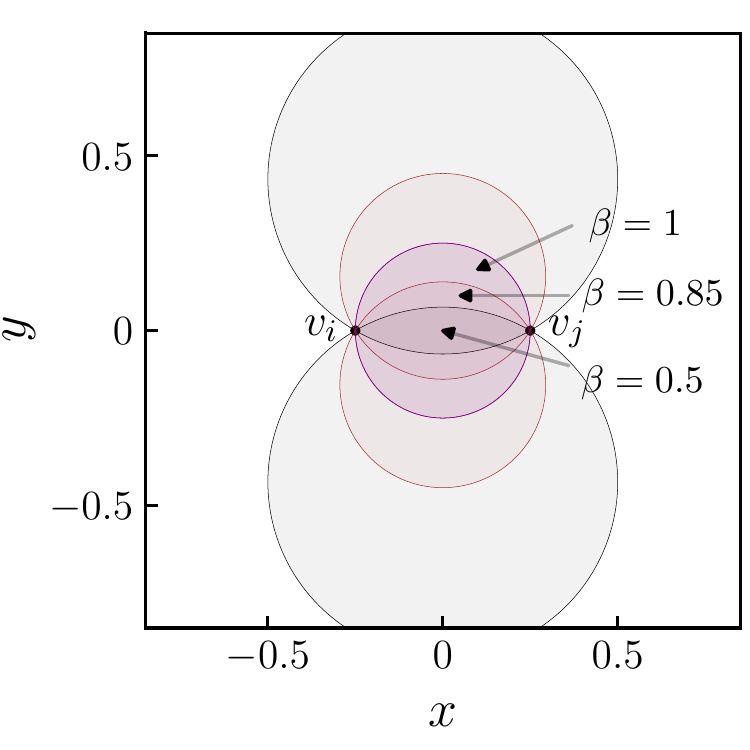}
\caption{Examples of neighborhoods ${\cal N}(v_i, v_j, \beta)$ for the vertices $v_i=(-0.25,0)$ and $v_j=(0.25,0)$ and 
several values of $\beta\le 1$.}
\label{Fig1}
\end{figure}

(1) {\it
for $0 < \beta \leq 1$, ${\cal N}(v_i, v_j, \beta)$ is the intersection of two discs, each with radius 
\begin{equation}
r = \frac{d(v_i, v_j)}{2\beta} \ ,
\end{equation}
having the segment $v_i v_j$ as a chord. The disc centers are located at
\begin{equation}\label{betaLeq1}
c^{-}_{+} = \frac{v_i + v_j}{2} \mp \frac{R(\pi/2)(v_j - v_i)}{2\beta}(1- \beta^2)^{1/2} ,
\end{equation}
where $R(\cdot)$ is a rotation matrix and $v_i$ and $v_j$ are the coordinate vectors of the corresponding vertices, 
namely}
\begin{equation}
R(\pi/2) = \left( \begin{array}{cc} 
0 & -1  \\ 
1 &0  \end{array} \right), \ v_i \equiv \left( \begin{array}{cc} 
x_i  \\ 
y_i  \end{array}\right), \ v_j \equiv \left( \begin{array}{cc} 
x_j  \\ 
y_j  \end{array}\right).
\end{equation} 
In Fig.~\ref{Fig1} we show some examples of neighborhoods ${\cal N}(v_i, v_j, \beta)$. 
We stress that in the limit $\beta\rightarrow0$ the neighborhood ${\cal N}$ becomes the straight line joining 
the vertices $v_i$ and $v_j$, so the network becomes fully connected.

(2) {\it
for $\beta > 1$ there are two proximity rules:}

\begin{figure}
    \centering
    \begin{minipage}{0.017\textwidth}
    \vspace{-3cm}(a)
    \end{minipage}
    \begin{minipage}{0.21\textwidth}
        \includegraphics[width=\linewidth]{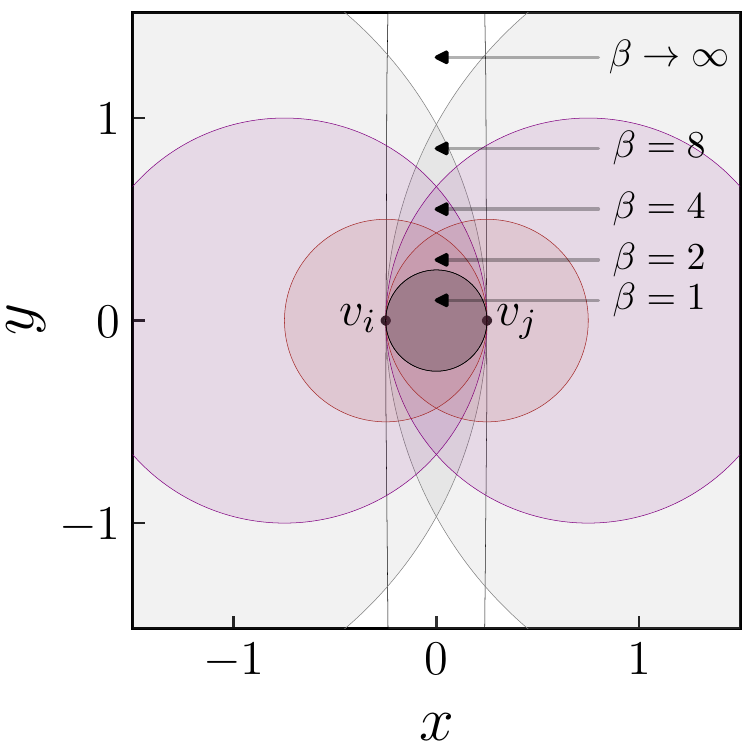}
    \end{minipage}%
     \begin{minipage}{0.017\textwidth}
     \vspace{-3cm}(b)
    \end{minipage}
    \begin{minipage}{0.21\textwidth}
        \includegraphics[width=\linewidth]{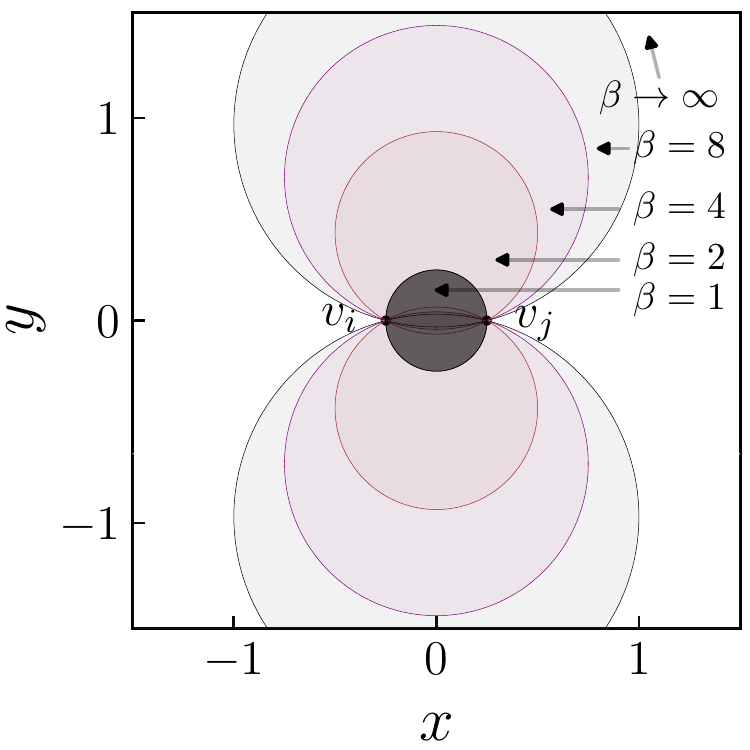}
    \end{minipage}
    \caption{Examples of (a) lune-based and (b) circle-based neighborhoods ${\cal N}(v_i, v_j, \beta)$ for the vertices 
$v_i=(-0.25,0)$ and $v_j=(0.25,0)$ and some values of $\beta\ge 1$.}
\label{Fig2}
\end{figure}

(2a) {\it
\textbf{lune-based BSG.} Here ${\cal N}(v_i, v_j, \beta)$ is the intersection of two discs, each with radius 
\begin{equation}
r= \frac{\beta d(v_i, v_j)}{2},
\label{rad}
\end{equation} 
whose centers are at}
\begin{align}\label{betaGeq1Lune}
c_1 &= \frac{\beta}{2} v_i + \left(1- \frac{\beta}{2}\right)v_j,\\
c_2 &= \frac{\beta}{2} v_j + \left(1- \frac{\beta}{2}\right)v_i .
\end{align}
In Fig. \ref{Fig2}(a) we can see the lunes of influence for different values of $\beta \geq1$. Note that in the limiting case $\beta \rightarrow\infty$, ${\cal N}(v_i, v_j, \beta)$ is an infinite strip of width $|v_i-v_j|$; thus, even for very large values of $\beta$ some connections may exist.

(2b) {\it
\textbf{circle-based BSG.} Here ${\cal N}(v_i, v_j, \beta)$ is the union of two discs, with radius given by
Eq.~(\ref{rad}), that pass through both $v_i$ and $v_j$. The disc centers are located at}
\begin{equation}\label{betaGeq1Circle}
c^{-}_{+} = \frac{v_i + v_j}{2} \mp \frac{R(\pi/2)(v_j - v_i)}{2}(\beta^2 - 1)^{1/2} .
\end{equation}
In Fig. \ref{Fig2}(b) we can see the circles of influence for different values of $\beta \geq1$. Note that in the limiting case $\beta \rightarrow\infty$, ${\cal N}(v_i, v_j, \beta)$ is the entire plane; therefore, for large enough values of $\beta$ the skeleton graph becomes a disconnected graph.

It is worth mentioning that for $\beta = 1$, Eqs.~(\ref{betaLeq1}), (\ref{betaGeq1Lune}) and (\ref{betaGeq1Circle}) 
reduce to the same expression. Indeed, the case $\beta=1$ is well known in the literature as Gabriel graph 
\cite{KRRRS} and addressed as a $1$-skeleton graph. Another well known case is $\beta = 2$, which is known as
relative neighborhood graph \cite{GTT}, in the lune-based formulation, and typically addressed as $2$-skeleton graph.


\section{Random $\beta$-skeleton graphs in the unit square}

\begin{figure}
    \centering
    \begin{minipage}{0.017\textwidth}
    \vspace{-3cm}(a)
    \end{minipage}
    \begin{minipage}{0.21\textwidth}
        \includegraphics[width=\linewidth]{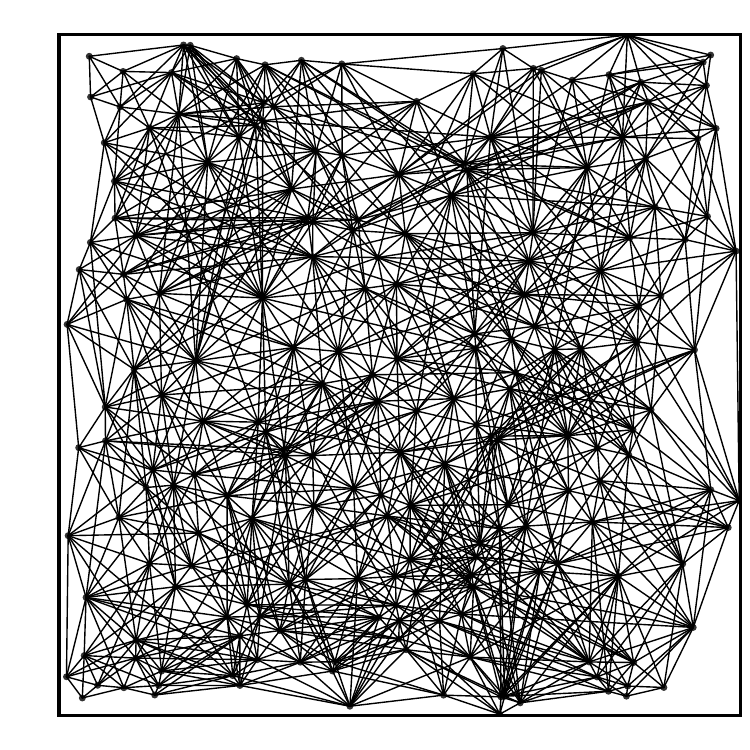}
    \end{minipage}%
     \begin{minipage}{0.017\textwidth}
     \vspace{-3cm}(b)
    \end{minipage}
    \begin{minipage}{0.21\textwidth}
        \includegraphics[width=\linewidth]{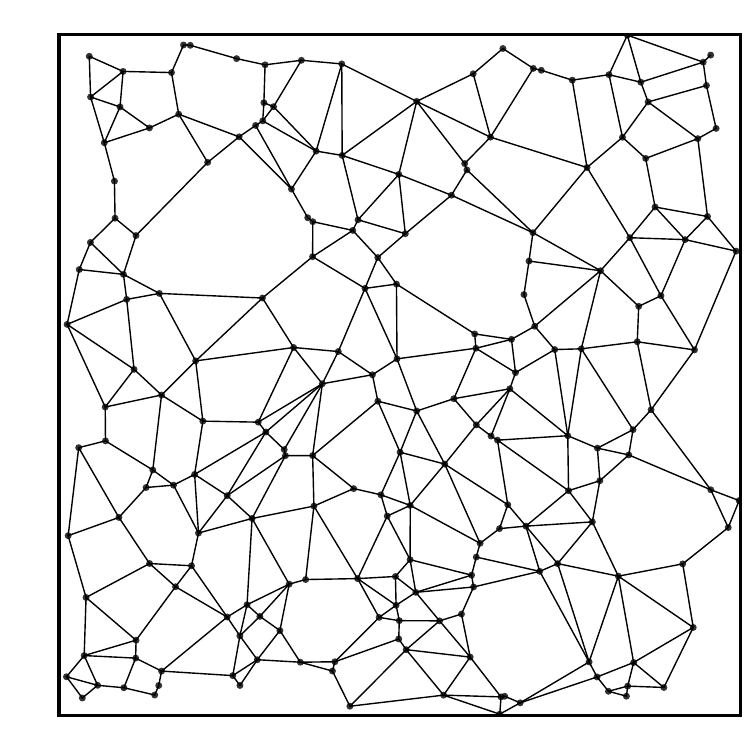}
    \end{minipage}
    \caption{BSGs with (a) $\beta=0.5$ and (b) $\beta=1$ for the same set of $N=200$ randomly 
distributed vertices in the unit square.}
\label{Fig3}
\end{figure}

In this work we consider randomly and independently distributed vertices in the unit square. As examples, in
Fig.~\ref{Fig3} we show BSGs with $\beta=0.5$ and $\beta=1$ for $N=200$. Note that we have used 
the same set of randomly distributed vertices in both panels. Here, since $\beta\le 1$ the proximity rule is unique.
Then, in Fig.~\ref{Fig4} we present BSGs for $\beta=1.5$ and $\beta=2$. There we consider both the 
lune-based (left panels) and the circle-based (right panels) proximity rules. We have used the same set of vertices 
of Fig.~\ref{Fig3}. From this figure it is clear that different proximity rules produce quite different networks. In 
particular, for a fixed value of $\beta$, lune-based skeleton graphs show higher connectivity than circle-based skeleton graphs.
We will characterize this feature by the use of geometrical and spectral properties below.

\subsection{Average degree}

A well known topological measure in graph theory is the degree of a vertex $k$, which is the number of 
edges incident to a given vertex. Here, since we are interested in random BSGs, we will consider 
the ensemble average degree $\langle k \rangle$ that we compute by averaging over all vertices of BSGs
with fixed parameter pairs $(N,\beta)$. 

On the one hand, in Figs.~\ref{Fig5} we plot $\langle k \rangle$ as a function of $N$ for random BSGs  
with several values of $\beta<1$ (i.e., when only one proximity rule applies). We observe that for fixed $\beta$, 
$\langle k \rangle$ increases for increasing $N$. Moreover, for fixed $N$, $\langle k \rangle$ increases for 
decreasing $\beta$; this confirms the expected scenario of completely connected networks in the limit
$\beta\to 0$.

On the other hand, in Figs.~\ref{Fig6} and~\ref{Fig7} we also plot $\langle k \rangle$ as a function of $N$ 
but now for random BSGs with $\beta \geq1$. We consider both lune-based (left panels) and
circle-based (right panels) proximity rules. For clarity, we group the data in the regimes $1\le\beta<2$ 
(Fig.~\ref{Fig6}) and $\beta\ge 2$ (Fig.~\ref{Fig7}).
First, let us concentrate on the BSGs constructed with the lune-based proximity rule, see left 
panels in Figs.~\ref{Fig6} and~\ref{Fig7}. There, we observe three different behaviors for $\langle k \rangle$:
(i) when $\beta$ is small, $\beta<2$, $\langle k \rangle$ is an increasing function of $N$; 
(ii) for intermediate values of $\beta$, $2\le \beta \le 20$, $\langle k \rangle$ is approximately constant 
for the values of $N$ we used in this work; and
(iii) when $\beta$ is large, $\beta>20$, $\langle k \rangle$ is a decreasing function of $N$.
This panorama is also observed for BSGs constructed with the circle-based proximity rule (see 
right panels in Figs.~\ref{Fig6} and~\ref{Fig7}) however shifted to smaller values of $\beta$; that is, 
$\langle k \rangle$ is approximately constant as a function of $N$ for $1.5\le \beta \le 4$. 


\begin{figure}
    \centering
    \begin{minipage}{0.017\textwidth}
    \vspace{-3cm}(a)
    \end{minipage}
    \begin{minipage}{0.21\textwidth}
    \textbf{Lune-based BSGs}
        \includegraphics[width=\linewidth]{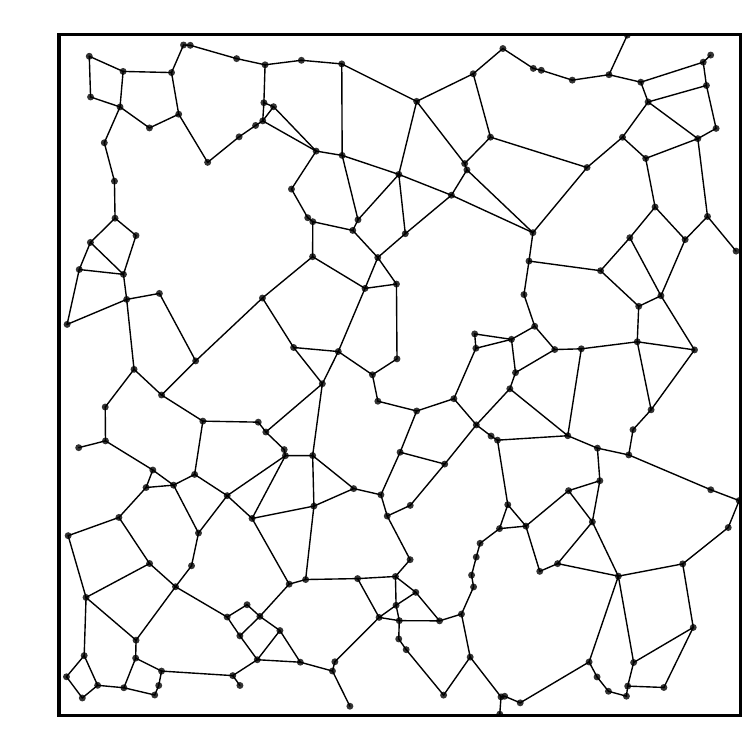}
    \end{minipage}%
     \begin{minipage}{0.017\textwidth}
     \vspace{-3cm}(b)
    \end{minipage}
    \begin{minipage}{0.21\textwidth}
    \textbf{Circle-based BSGs}
        \includegraphics[width=\linewidth]{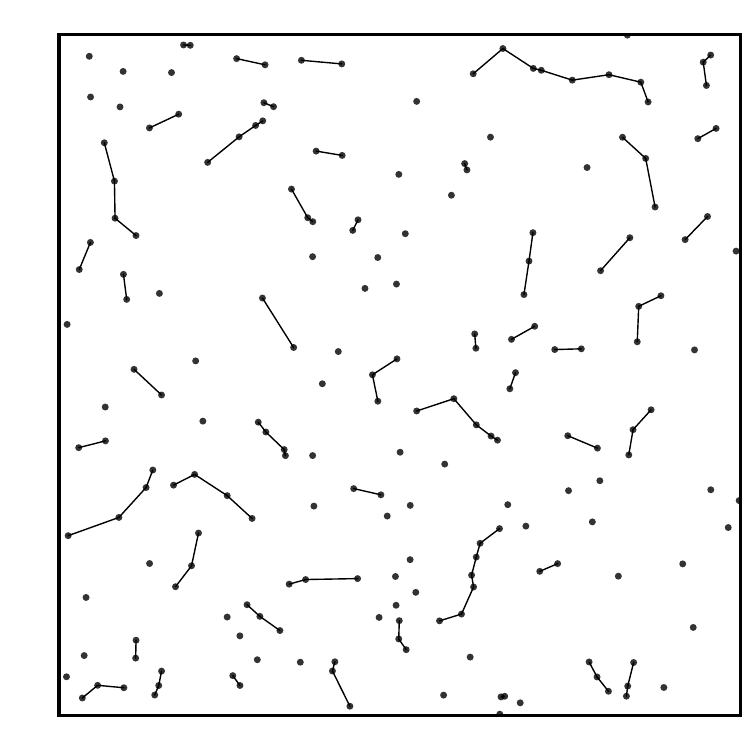}
    \end{minipage}
     \begin{minipage}{0.017\textwidth}
    \vspace{-3cm}(c)
    \end{minipage}
    \begin{minipage}{0.21\textwidth}
        \includegraphics[width=\linewidth]{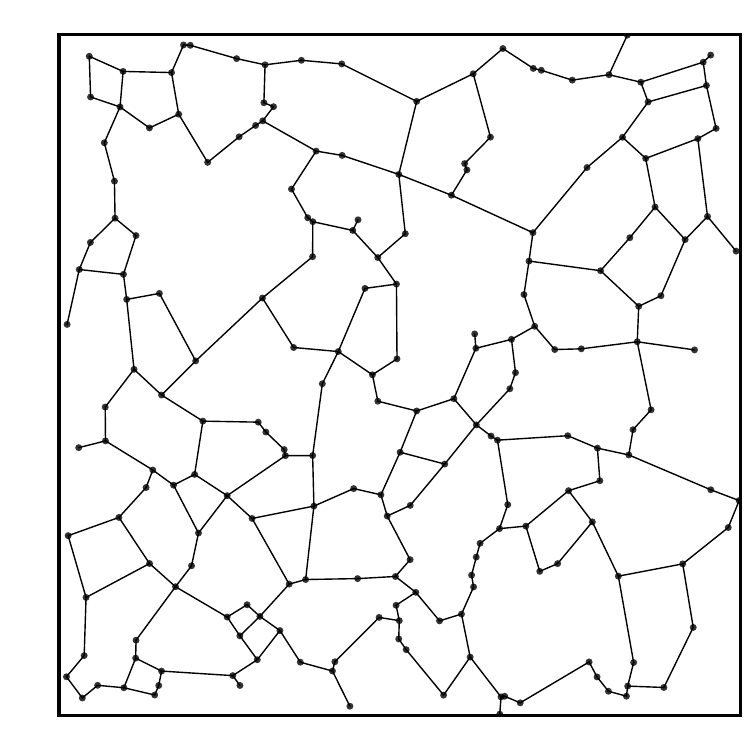}
    \end{minipage}%
     \begin{minipage}{0.017\textwidth}
     \vspace{-3cm}(d)
    \end{minipage}
    \begin{minipage}{0.21\textwidth}
        \includegraphics[width=\linewidth]{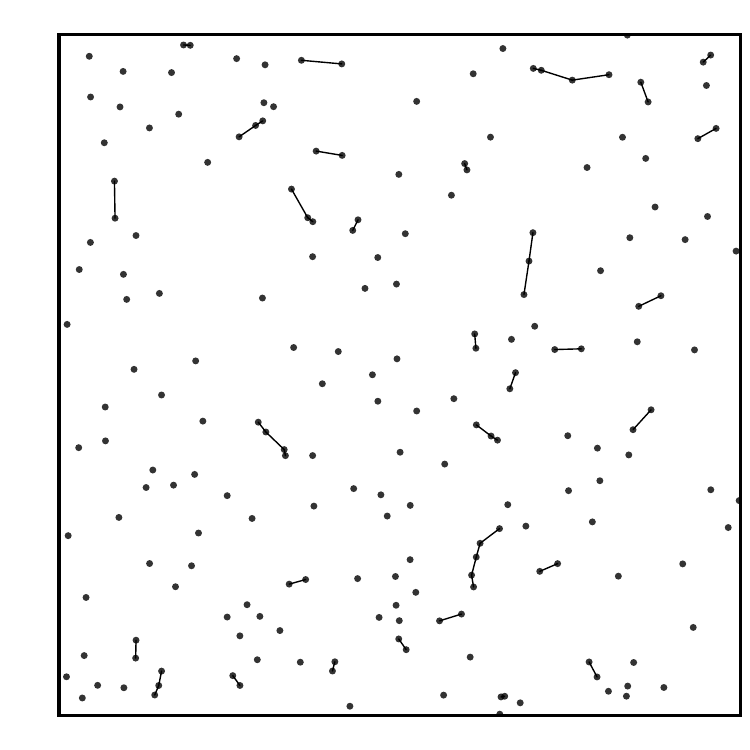}
    \end{minipage}
    \caption{BSGs with (a,b) $\beta=1.5$ and (c,d) $\beta=2$ for the same set of randomly distributed 
vertices of Fig.~\ref{Fig3}. The lune-based [circle-based] proximity rule was used in the left [right] panels.}
\label{Fig4}
\end{figure}

\begin{figure}
\includegraphics[width=.45\textwidth]{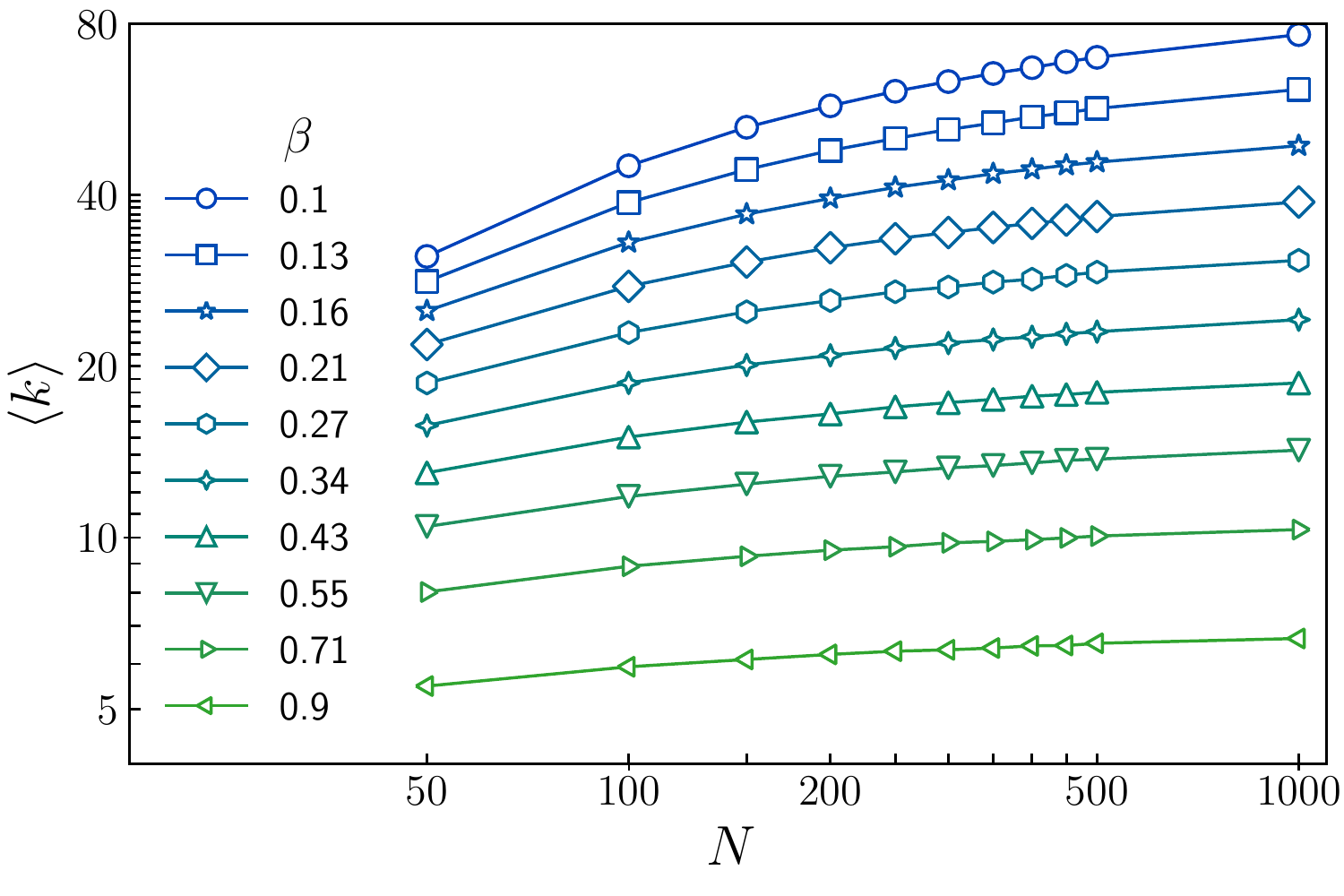}
\caption{Average degree $\langle k \rangle$ as a function of $N$ for random BSGs with
$\beta<1$. Here, the standard deviation error is smaller than the symbol size.}
\label{Fig5}
\end{figure}


\begin{figure*}
    \centering
    \begin{minipage}{0.017\textwidth}
    \vspace{-4cm}(a)
    \end{minipage}
    \begin{minipage}{0.45\textwidth}
        \textbf{ Lune-based proximity rule}\par
        \medskip
        \includegraphics[width=\linewidth, valign=t]{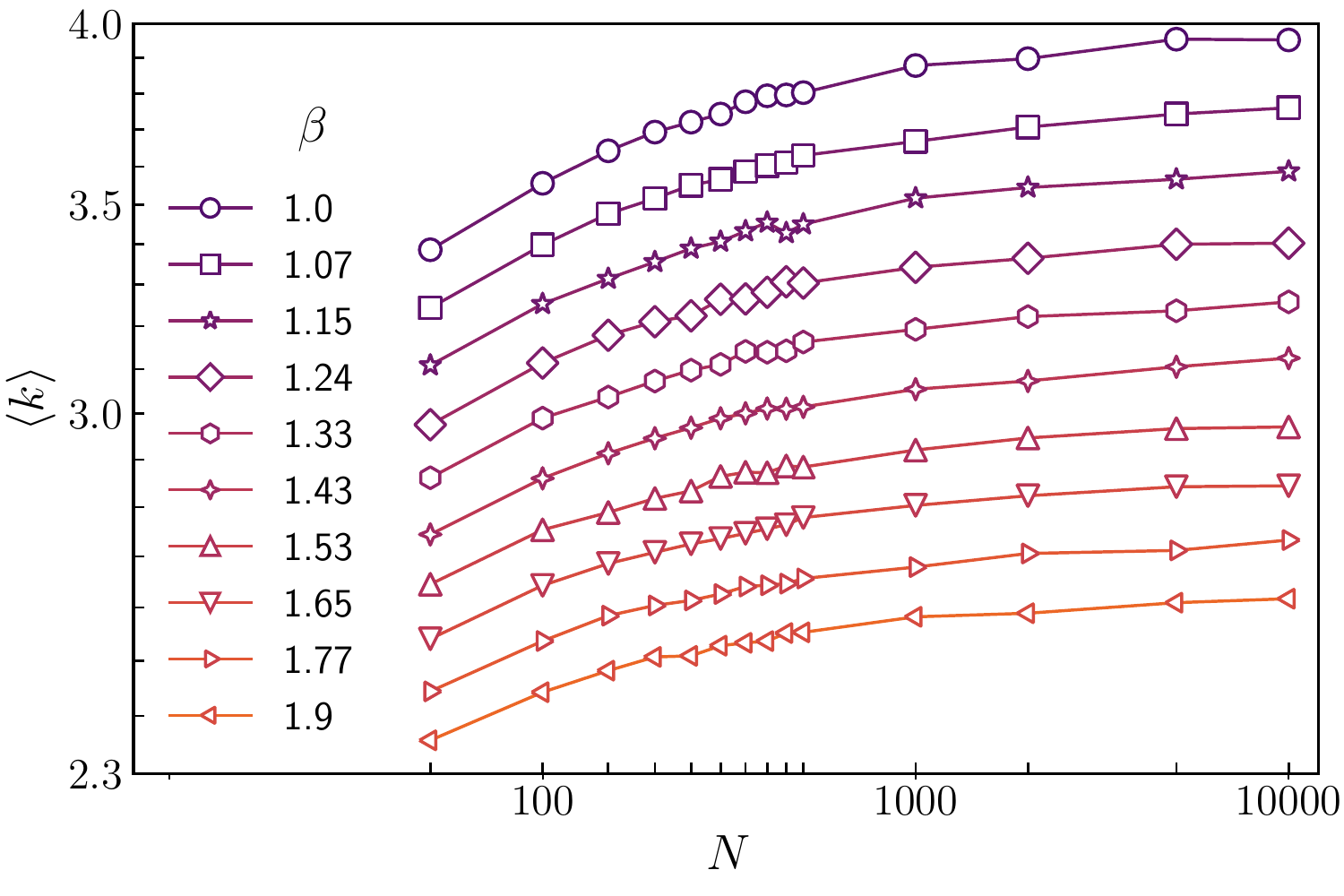}
    \end{minipage}%
     \begin{minipage}{0.017\textwidth}
     \vspace{-4cm}(b)
    \end{minipage}
    \begin{minipage}{0.45\textwidth}
        \textbf{Circle-based proximity rule}\par
        \medskip
        \includegraphics[width=\linewidth, valign=t]{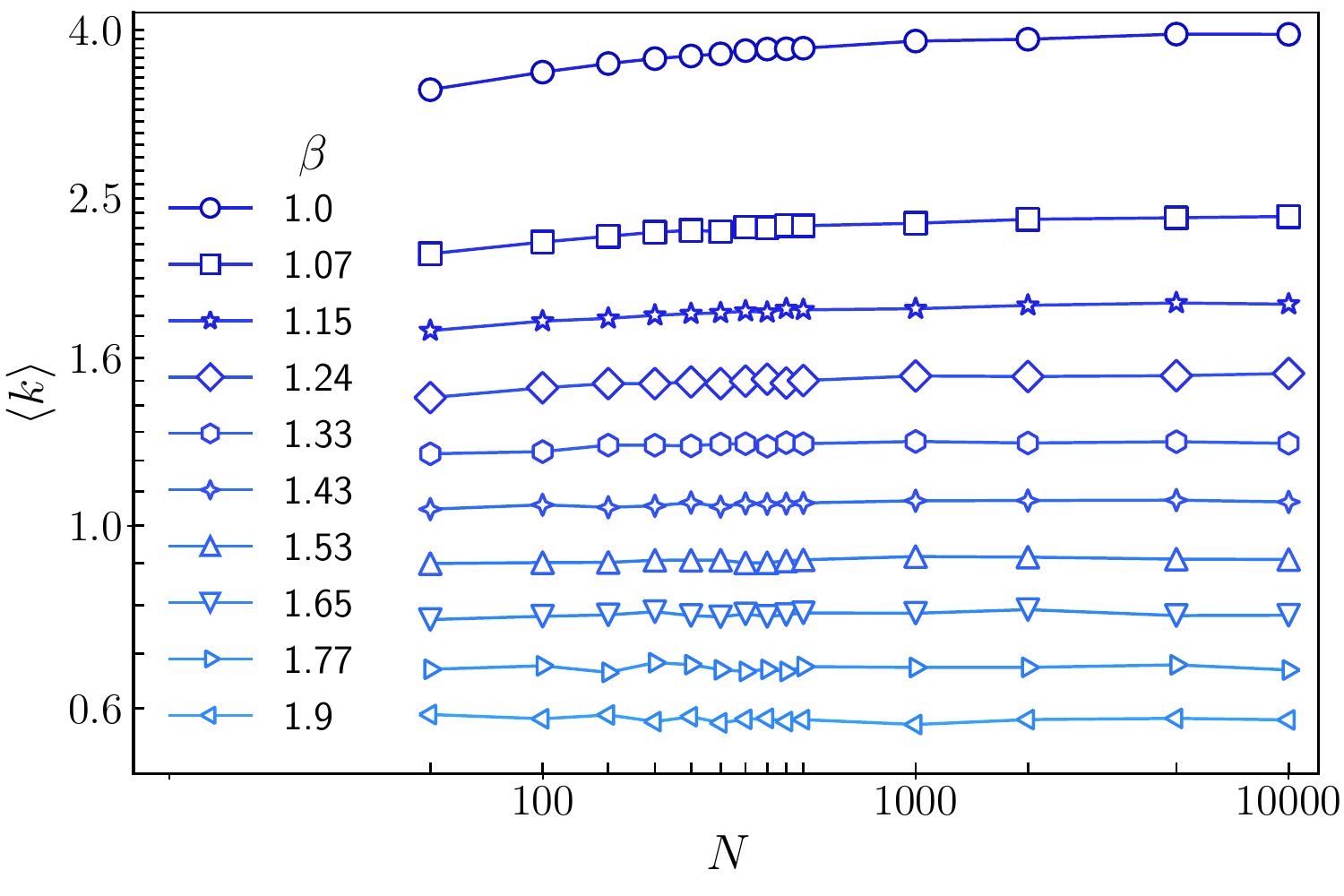}
    \end{minipage}
    \caption{Average degree $\langle k \rangle$ as a function of $N$ for random BSGs with
$1\le\beta<2$. In both panels the standard deviation error is smaller than the symbol size.}
\label{Fig6}
\end{figure*}


\begin{figure*}
    \centering
    \begin{minipage}{0.017\textwidth}
    \vspace{-4cm}(a)
    \end{minipage}
    \begin{minipage}{0.45\textwidth}
        \textbf{ Lune-based proximity rule}\par
        \medskip
        \includegraphics[width=\linewidth, valign=t]{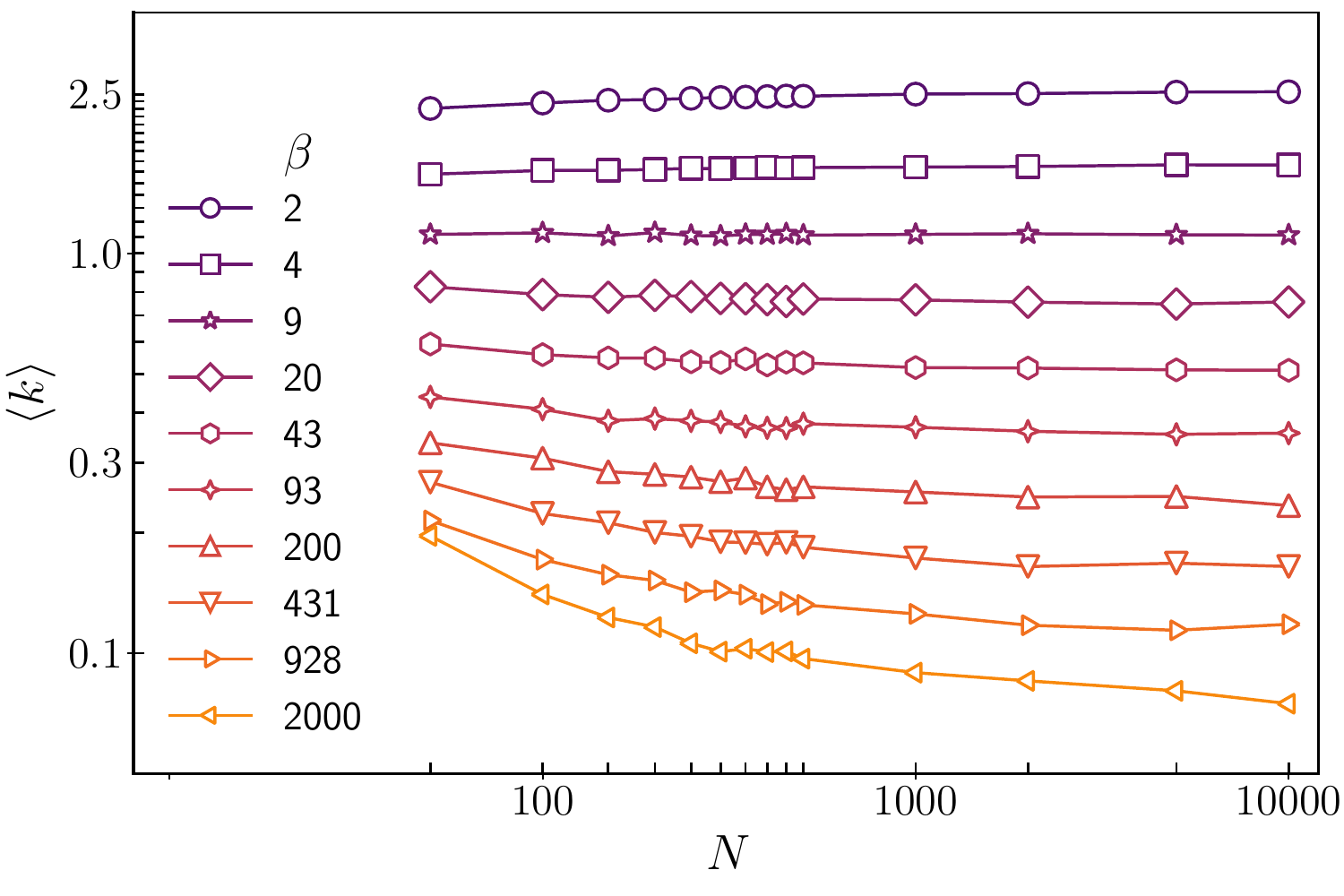}
    \end{minipage}%
     \begin{minipage}{0.017\textwidth}
     \vspace{-4cm}(b)
    \end{minipage}
    \begin{minipage}{0.45\textwidth}
        \textbf{Circle-based proximity rule}\par
        \medskip
        \includegraphics[width=\linewidth, valign=t]{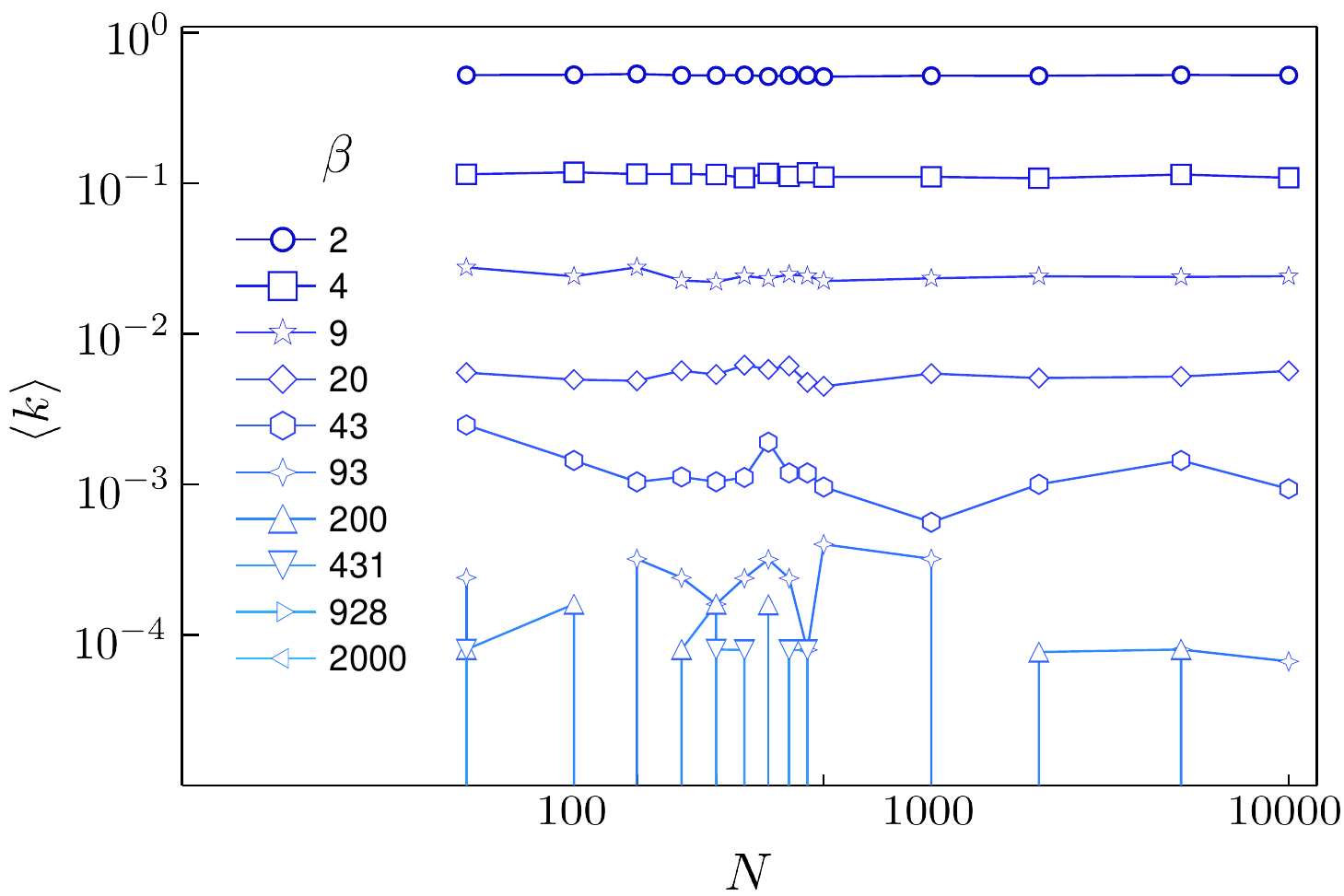}
    \end{minipage}
    \caption{Average degree $\langle k \rangle$ as a function of $N$ for random BSGs with
$\beta\ge 2$. In (a) the standard deviation error is smaller than the symbol size.}
\label{Fig7}
\end{figure*}

From the observations above we can concluded that for intermediate values of $\beta$ (including relative 
neighbor graphs) the BSGs are very stable graphs in the sense that the average degree remains 
constant even in the presence of strong vertex density fluctuations. 

The main difference we can observe between random BSGs constructed with the lune-based 
and circle-based proximity rules is that in the circle-based case the networks become disconnected for relatively
smaller values of $\beta$ than in the lune-based case; compare Figs.~\ref{Fig7}(a) and~\ref{Fig7}(b) and also
left and right panels in Fig.~\ref{Fig4}.
Indeed, from Fig.~\ref{Fig7}(b) it is clear that when $\beta \ge 200$ it is highly probable to have completely disconnected networks.

\section{Weighted random $\beta$-skeleton graphs}

Here, in order to use Random Matrix Theory (RMT) results as a reference, we include weights, particularly 
random weights, to the random BSGs defined above.
Specifically, we choose the non-vanishing elements of the corresponding adjacency matrices ${\bf A}$ to 
be statistically independent random variables drawn from a normal distribution with zero mean 
$\langle A_{ij} \rangle=0$ and variance $\langle |A_{ij}|^2 \rangle=(1+\delta_{ij})/2$, where $\delta_{ij}$ is 
the Kronecker delta. Therefore, a diagonal adjacency random matrix is obtained for isolated vertices
(known in RMT as the Poisson case), whereas the Gaussian Orthogonal 
Ensemble (GOE) is recovered when the graph is fully connected.

Below we use exact numerical diagonalization to obtain the eigenvalues $\lambda^m$ and eigenvectors
$\Psi^m$ ($m=1\ldots N$) of the adjacency matrices of large ensembles of weighted random BSGs 
characterized by $\beta$ and $N$.

\subsection{Spectral properties}

\begin{figure}
    \centering
    \begin{minipage}{0.017\textwidth}
    \vspace{-4.5cm}(a)
    \end{minipage}
    \begin{minipage}{0.45\textwidth}
        \includegraphics[width=\linewidth, valign=t]{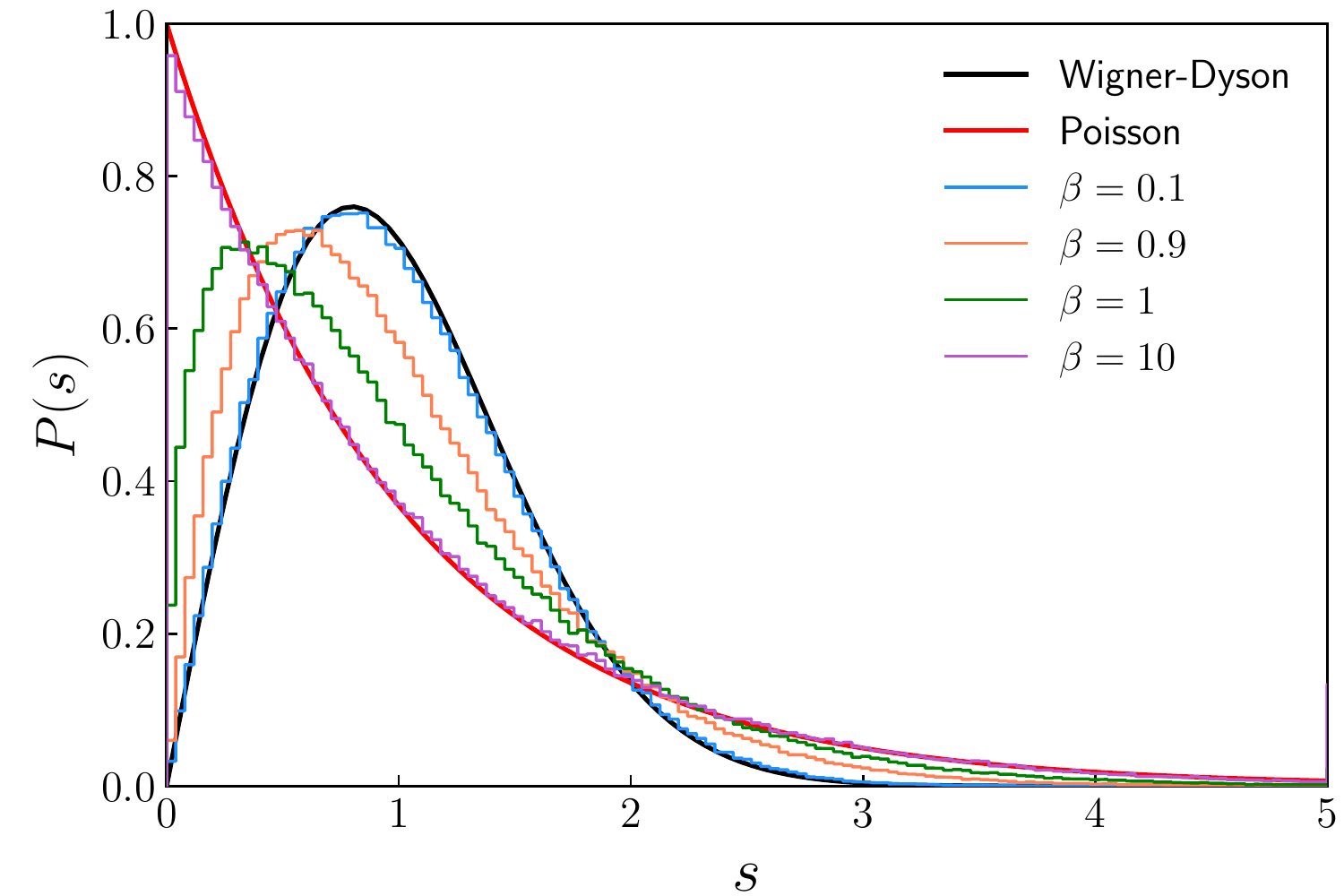}
    \end{minipage}%
    
      \begin{minipage}{0.017\textwidth}
     \vspace{-4.5cm}(b)
    \end{minipage}
    \begin{minipage}{0.45\textwidth}
        \includegraphics[width=\linewidth, valign=t]{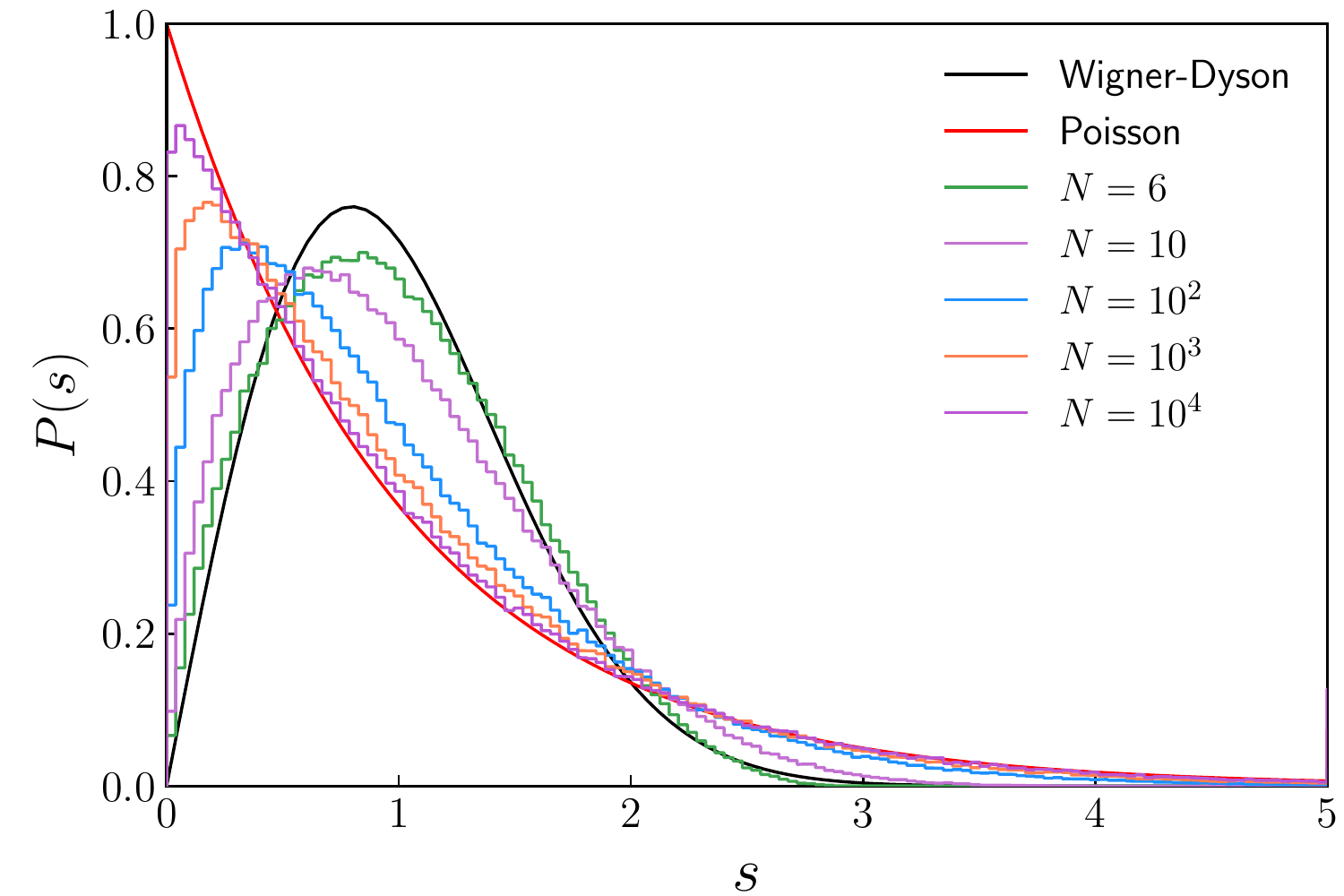}
    \end{minipage}
    
    \caption{Nearest-neighbor energy level spacing distribution $P(s)$ for weighted random BSGs with 
(a) [(b)] $N = 10^2$ [$\beta=1$] and several values of $\beta$ [$N$]. Here we use the lune-based 
proximity rule to construct the BSGs. Full lines correspond to Poisson and Wigner-Dyson distribution 
functions given by Eqs.~(\ref{P}) and~(\ref{WD}), respectively.}
\label{Fig8}
\end{figure}

In order to characterize the spectra of weighted random BSGs, we use the nearest-neighbor 
energy level spacing distribution $P(s)$~\cite{metha}; a widely used tool in RMT. For $\beta\to\infty$, 
i.e., when the vertices 
in the random BSGs are mostly isolated, the corresponding adjacency matrices are almost 
diagonal and, regardless of the size of the graph, $P(s)$ should be close to the exponential distribution,
\begin{equation}
\label{P}
P(s) = \exp(-s) \ ,
\end{equation}
which is better known in RMT as Poisson distribution. In the opposite limit, $\beta\to 0$, when the weighted 
BSGs are fully connected, the adjacency matrices become members of the GOE (real and  
symmetric full random matrices) and $P(s)$ closely follows the Wigner-Dyson distribution,
\begin{equation}
\label{WD}
P(s) = \frac{\pi}{2} s \exp \left(- \frac{\pi}{4} s^2 \right) \ .
\end{equation}
Thus, for a fixed graph size $N$, by increasing $\beta$ from zero to infinity, the shape of $P(s)$ is 
expected to evolve from the Wigner-Dyson distribution to the Poisson distribution. Moreover, for a 
fixed value of $\beta$, the increase in the density of vertices $N$ also produces changes in the shape of
$P(s)$. In Fig.~\ref{Fig8} we explore both scenarios.


\begin{figure}
\includegraphics[width=.45\textwidth]{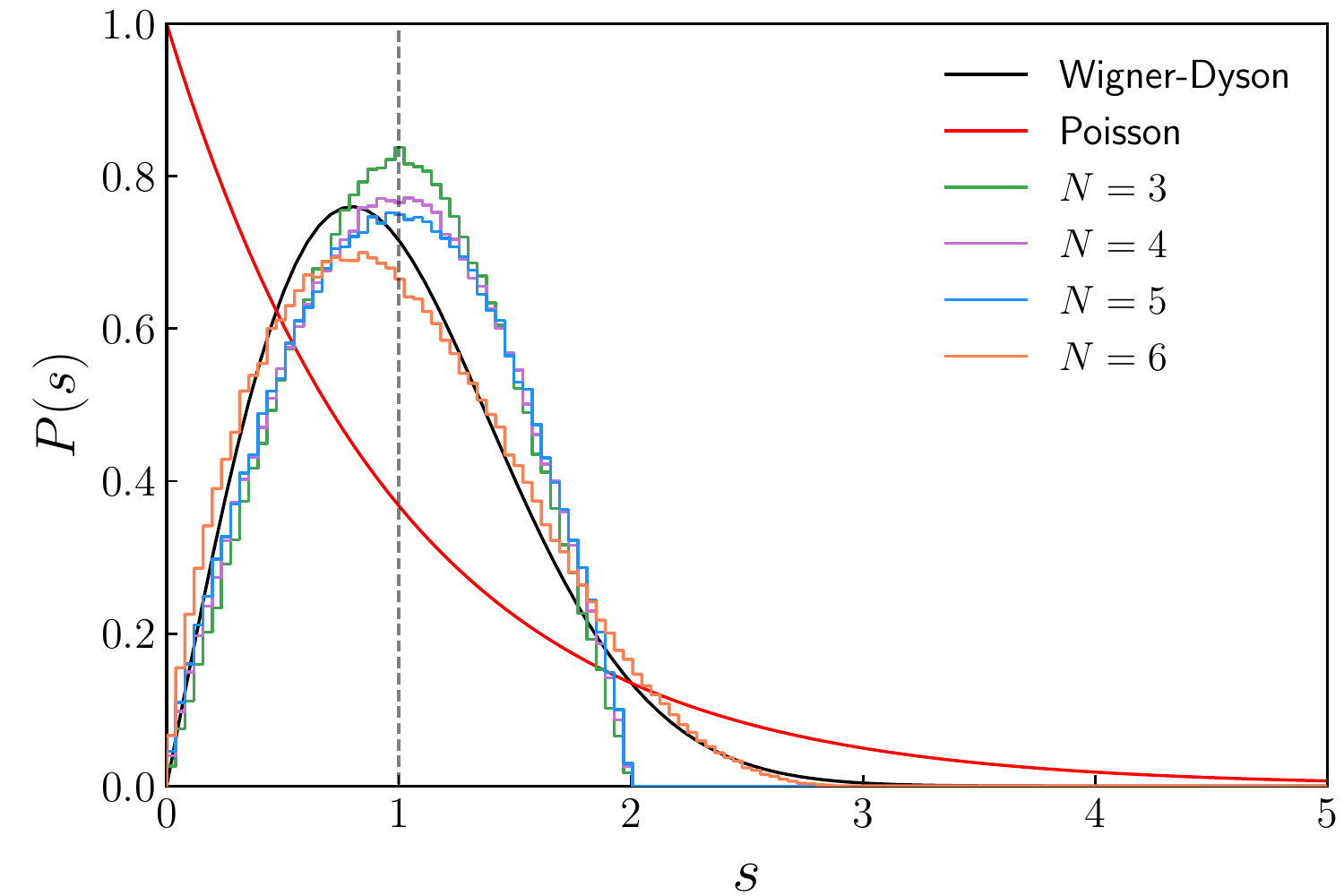}
\caption{Nearest-neighbor energy level spacing distribution $P(s)$ for weighted random BSGs with 
$\beta=1$ and $N\le 6$.  Vertical dashed line at $s=1$ is plotted to guide the eye.}
\label{Fig8.1}
\end{figure}

We construct histograms of $P(s)$ from $N/2$ unfolded spacings \cite{metha},
$s_m=(\lambda^{m+1}-\lambda^m)/\Delta$, around the band center of a large number of graph realizations
(such that all histograms are constructed with $5\times 10^5$ spacings). 
Here, $\Delta$ is the mean level spacing computed for each adjacency matrix.
Then, Fig.~\ref{Fig8} presents histograms of $P(s)$ for the adjacency matrices of weighted random 
BSGs:
In Fig.~\ref{Fig8}(a) the graph size is fixed to $N = 10^2$ and $\beta$ takes the values 0.1, 0.9, 1, and 10.
In this figure we observe a complete transition in the shape of $P(s)$ from the Wigner-Dyson to Poisson
distribution functions (also shown as reference) for increasing $\beta$.
In Fig.~\ref{Fig8}(b) the parameter $\beta$ is set to one (Gabriel graph) while $N$ increases from 6 to $10^4$.
Here, in contrast to Fig.~\ref{Fig8}(a), we do not observe a complete transition from Wigner-Dyson to Poisson 
in the shape of $P(s)$. From Fig.~\ref{Fig8}(b) one may expect that by decreasing further the number of vertices 
$N$ the Wigner-Dyson shape could emerge; however, this is not the case, as shown in Fig.~\ref{Fig8.1}.
There we observe that for $N<6$ the $P(s)$ becomes symmetric with respect to $s=1$. It is important to
stress that we have nor observed this shape for the $P(s)$ before.
Indeed, in other random network models embedded in the plane, such as random regular 
graphs and random rectangular graphs (RRGs) (for the definition and general properties of RRGs the reader is referred to~\cite{ES15}), we did observe the full transition from Wigner-Dyson to 
Poisson for the $P(s)$ as a function of the density of vertices for a fixed value of the proximity rule parameter~\cite{AMGM18}. 

Now, in order to characterize the shape of $P(s)$ for weighted random BSGs we use the Brody 
distribution~\cite{B73,B81} 
\begin{equation}
\label{B}
P(s) = (\mu +1) a_{\mu} s^{\mu}  \exp\left(- a_{\mu} s^{\mu+1}\right) \ ,
\end{equation}
where $a_{\mu} = \Gamma[(\mu+2)/(\mu+1)]^{\mu+1}$, $\Gamma(\cdot)$ is the gamma 
function, and $\mu$, known as Brody parameter, takes values in the range $[0,1]$.
Equation~(\ref{B}) was originally derived to provide an interpolation expression for $P(s)$ in the 
transition from Poisson to Wigner-Dyson distributions, serving as a measure for the degree of mixing 
between Poisson and GOE statistics. In fact, $\mu=0$ and $\mu=1$ in Eq.~(\ref{B})
produce Eqs.~(\ref{P}) and~(\ref{WD}), respectively. 
In particular, as we show below, the Brody parameter will allows us to identify the onset of the 
localization transition for random BSGs.
It is also relevant to mention that the Brody distribution has been applied to study other complex 
networks models, see e.g.~\cite{AMGM18,BJ07,JB08,ZYYL08,J09,JB07,MAM15,DGK16}.
In fact, we found that Eq.~(\ref{B}) provides excellent fittings to the histograms of $P(s)$ of weighted random 
BSGs. For example, the fittings to the histograms in Fig.~\ref{Fig8}(a) [Fig.~\ref{Fig8}(b)] 
(not shown to avoid figure saturation) provide: 
$\mu(\beta)=0.953(0.1)$, $0.624(0.9)$, $0.276(1)$, and $0.002(10)$ 
[$\mu(N)=0.872(6)$, $0.645(10)$, $0.316(10^2)$, $0.145(10^3)$, and $0.0468(10^4)$].
It is important to remark that for $N\leq 5$ the $P(s)$ can not be fitted by the Brody distribution, see 
Fig.~\ref{Fig8.1}, therefore we will not consider small graph sizes in our analysis below.

\begin{figure*}
    \centering
    \begin{minipage}{0.017\textwidth}
    \vspace{-4cm}(a)
    \end{minipage}
    \begin{minipage}{0.45\textwidth}
        \textbf{ Lune-based proximity rule}\par
        \medskip
        \includegraphics[width=\linewidth, valign=t]{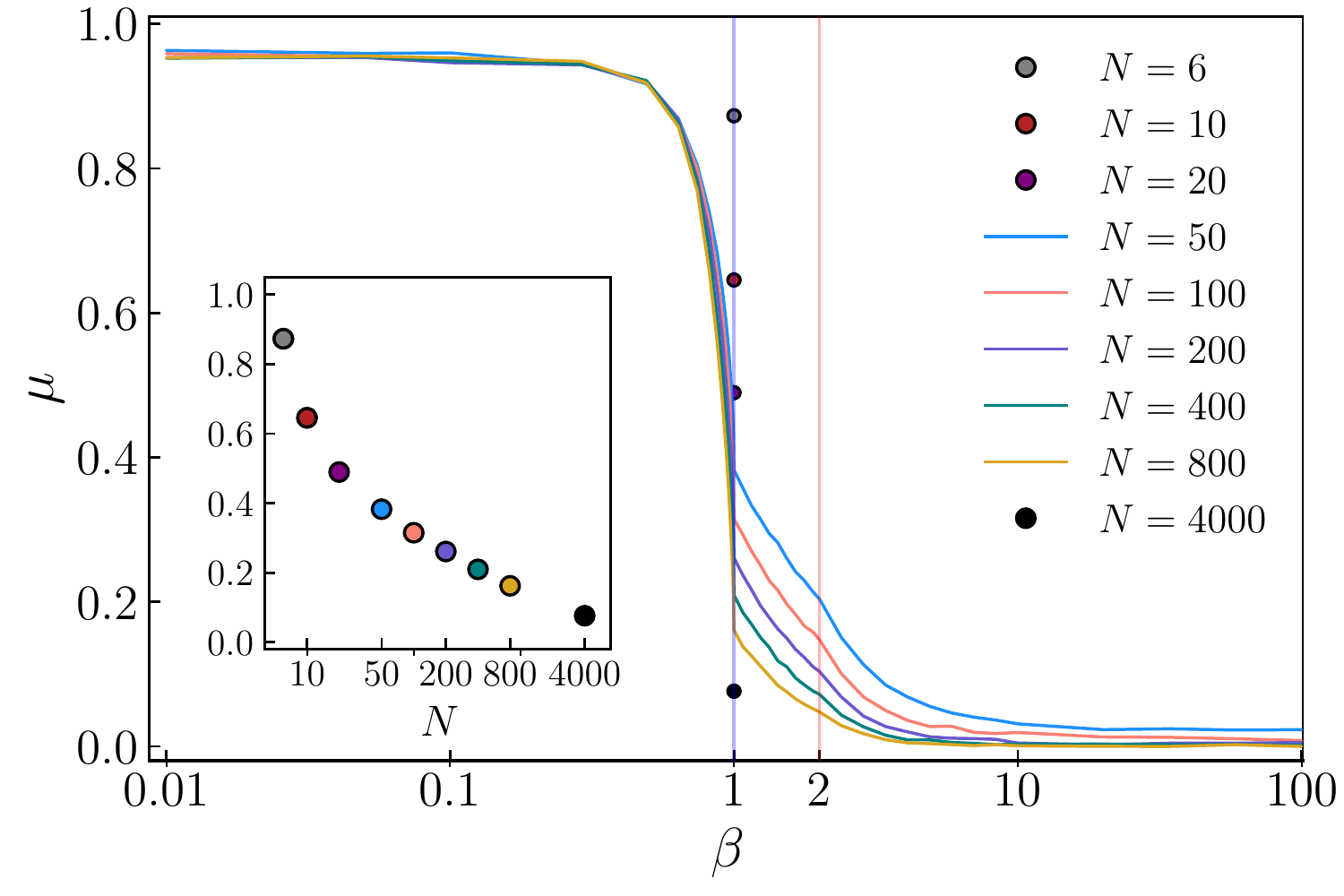}
    \end{minipage}%
     \begin{minipage}{0.017\textwidth}
     \vspace{-4cm}(b)
    \end{minipage}
    \begin{minipage}{0.45\textwidth}
        \textbf{Circle-based proximity rule}\par
        \medskip
        \includegraphics[width=\linewidth, valign=t]{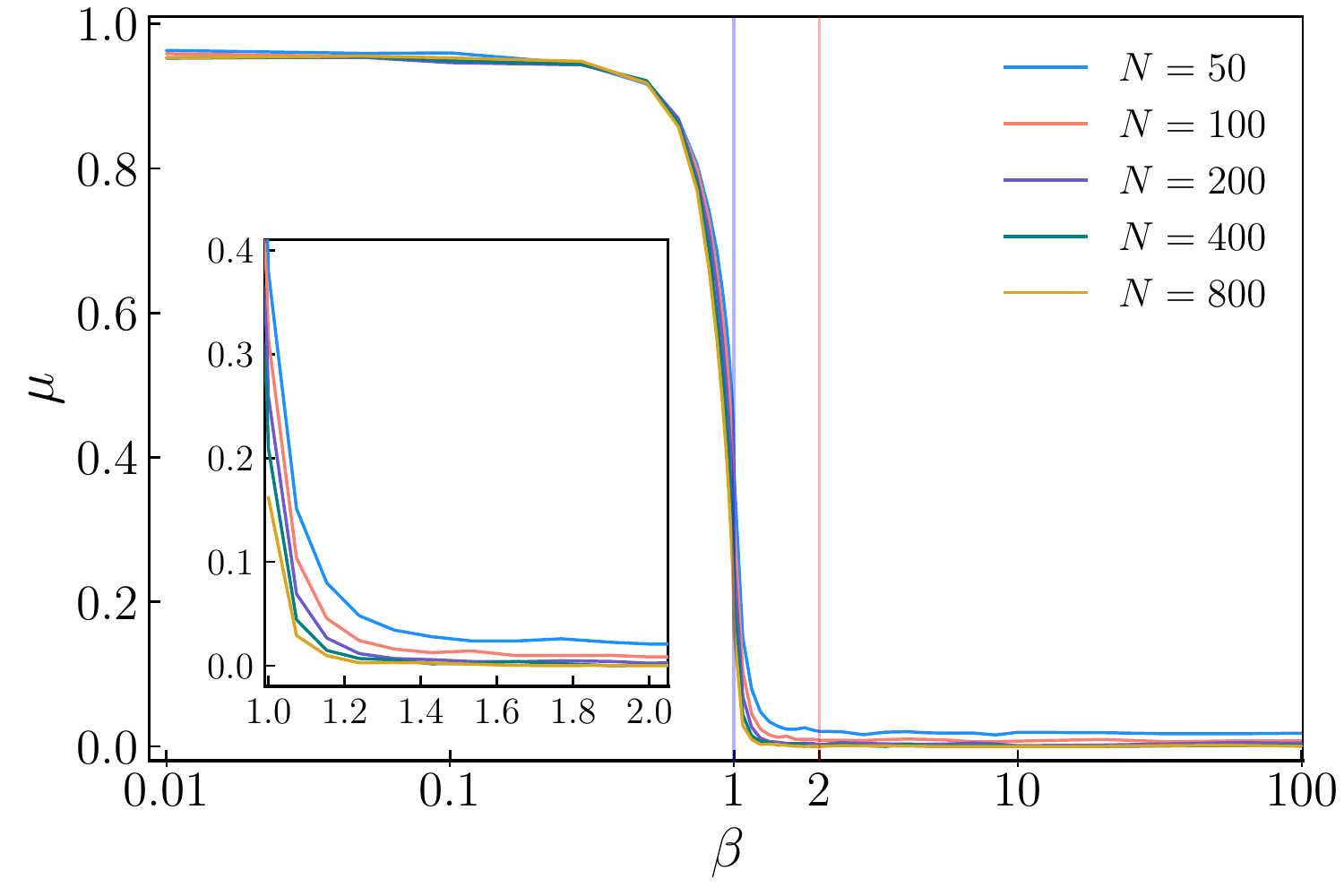}
    \end{minipage}
    \caption{Brody parameter $\mu$ as a function of $\beta$ for several values of $N$ for (a) lune-based and 
(b) circle-based weighted random BSGs. 
Vertical lines in the main panels indicate Gabriel graphs ($\beta=1$) and relative neighbor graphs ($\beta=2$).
The inset in (a) shows $\mu$ vs.~$N$ for $\beta=1$ (Gabriel graphs). The inset in (b) is an enlargement of 
the main panel in the interval $1\leq\beta \leq2$.}
\label{Fig9}
\end{figure*}

Thus we now perform a systematic study of the Brody parameter $\mu$ as a function of the parameters 
$\beta$ and $N$ of the BSGs. To this end, we construct histograms of $P(s)$ for a large number 
of parameter combinations to extract the corresponding values of $\mu$ by fitting them using 
Eq.~(\ref{B}). Figure~\ref{Fig9} reports $\mu$ versus $\beta$ for five different graph sizes for both proximity 
rules; lune-based and circle-based. Notice that in all cases the behavior of $\mu$ is similar for increasing 
$\beta$: for small $\beta$ (i.e.,~$\beta\le 0.1$) $\mu$ is approximately constant and equal to $0.96$;
then $\mu$ decreases fast for $\beta$ approaching one; 
and finally, for $\beta>1$, $\mu$ continues decreasing but slowly when $\beta$ is further increased.
For large $\beta$ (i.e.,~$\beta> 10$) and large $N$, $\mu\approx 0$.


Indeed, from Fig.~\ref{Fig9} we can conclude that our model of weighted random BSGs 
undergoes a clear and sharp transition at $\beta=1$ from a regime very close to the GOE regime (mostly 
connected vertices), $\mu\approx 0.96$, to the Poisson regime (mostly isolated vertices), $\mu\approx 0$, 
as a function of $\beta$. 
What is remarkable is that this delocalization-to-localization transition seems to be independent of the density 
of vertices $N$; in fact, the larger the value of $N$ the sharper the transition at $\beta=1$ is.
It is interesting to recall that we have also identified a delocalization-to-localization transition in random regular 
graphs as a function of the proximity rule parameter~\cite{AMGM18}; however, for that model the transition is 
rather smooth and importantly depends on the density of vertices.
In addition, in the inset of Fig.~\ref{Fig9}(a) we present the values of $\mu$ for increasing $N$ for Gabriel
graphs where we include the case $N=4000$.

The main difference we can observe between random BSGs constructed with the lune-based 
and circle-based proximity rules is that the Poisson limit is approached faster in the circle-based case, which 
was already expected from the analysis of the average degree of the previous Subsection since there it was 
shown that circle-based BSGs become completely disconnected for relatively smaller values of $\beta$.

\subsection{Eigenvector properties}

\begin{figure*}
    \centering
    \begin{minipage}{0.017\textwidth}
    \vspace{-4cm}(a)
    \end{minipage}
    \begin{minipage}{0.45\textwidth}
        \textbf{ Lune-based proximity rule}\par
        \medskip
        \includegraphics[width=\linewidth, valign=t]{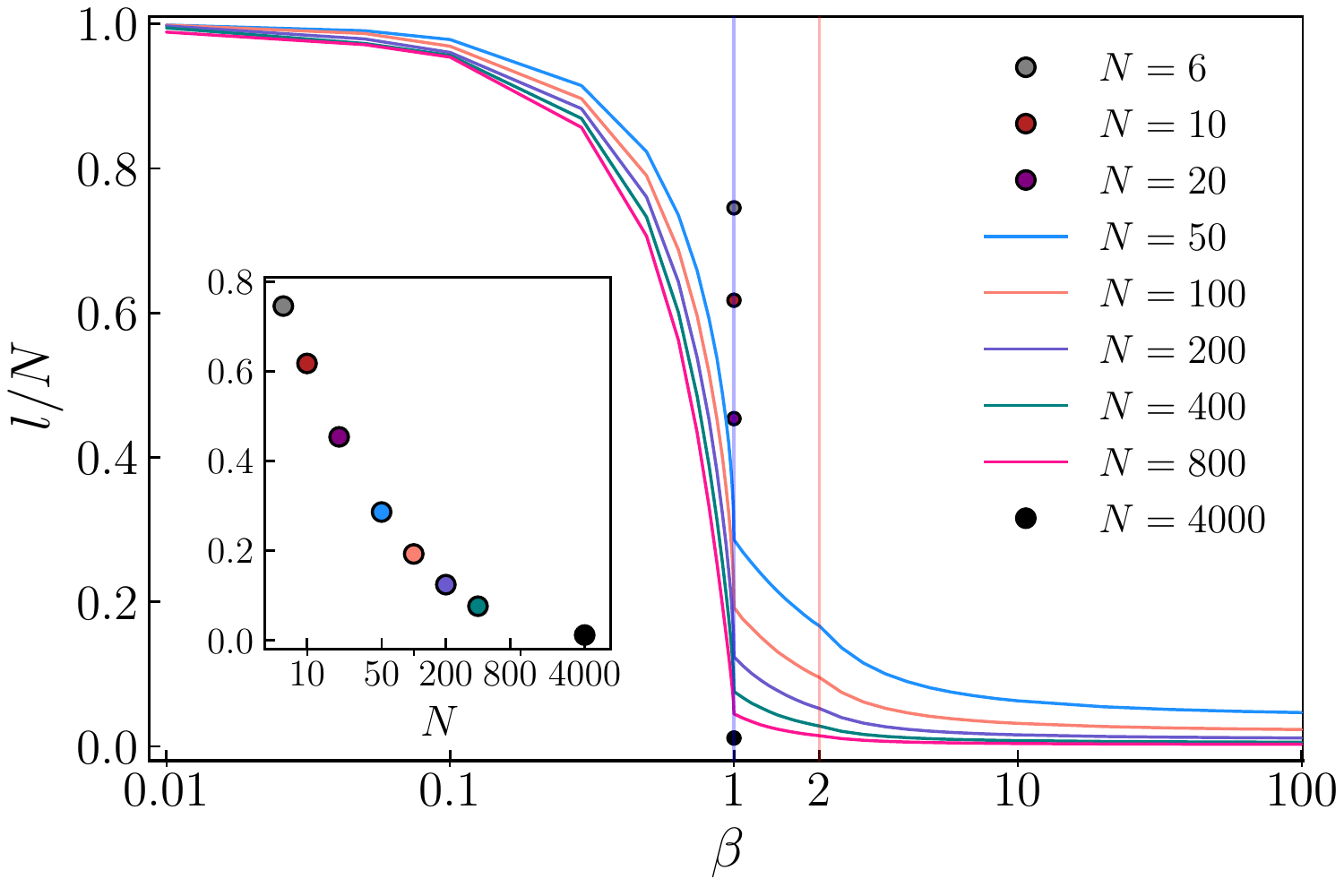}
    \end{minipage}%
     \begin{minipage}{0.017\textwidth}
     \vspace{-4cm}(b)
    \end{minipage}
    \begin{minipage}{0.45\textwidth}
        \textbf{Circle-based proximity rule}\par
        \medskip
        \includegraphics[width=\linewidth, valign=t]{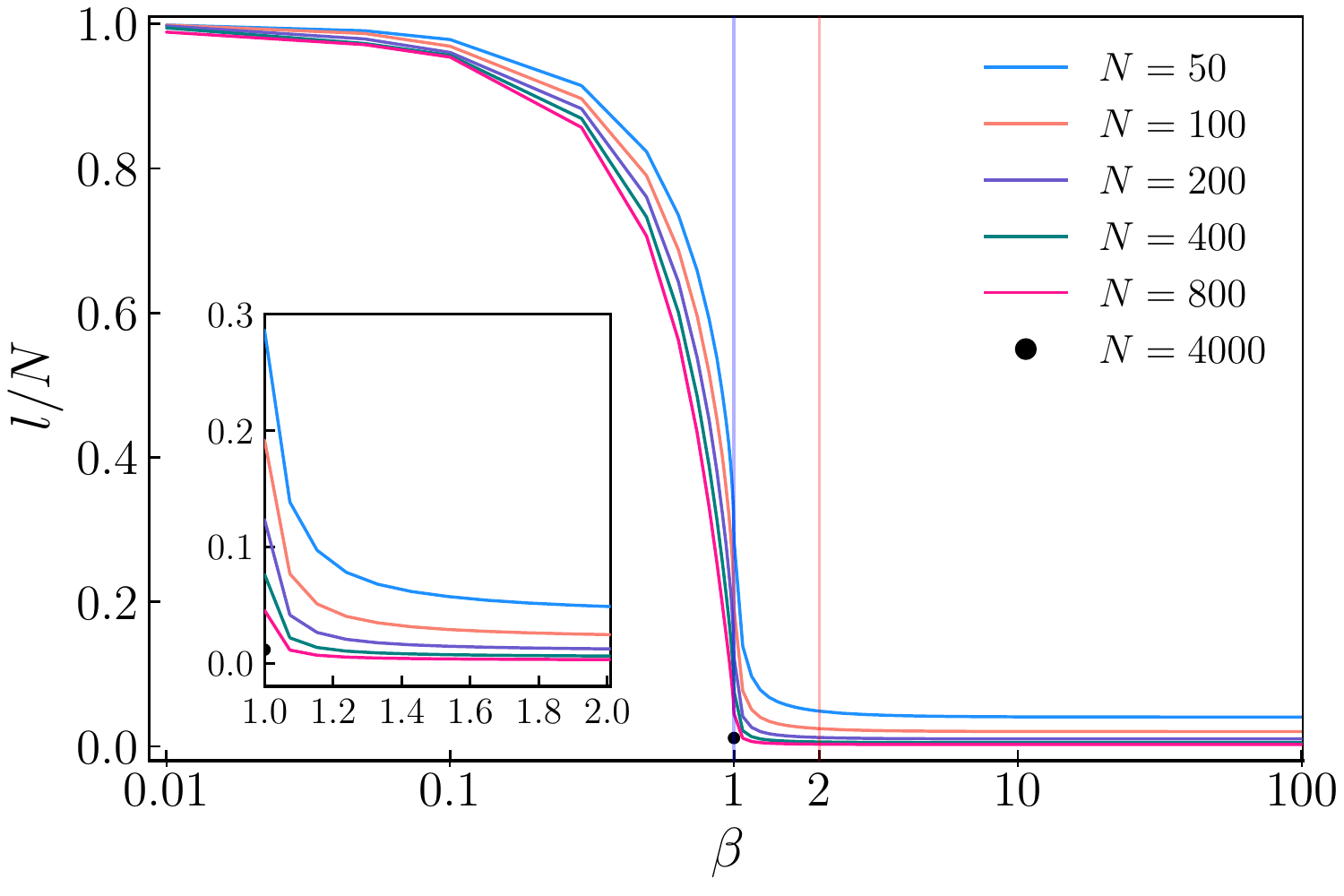}
    \end{minipage}
    \caption{Entropic eigenvector localization length $\ell$ (normalized to $N$) as a function of $\beta$ for 
several values of $N$ for (a) lune-based and  (b) circle-based weighted random BSGs. 
Vertical lines in the main panels indicate Gabriel graphs ($\beta=1$) and relative neighbor graphs ($\beta=2$).
The inset in (a) shows $\ell$ vs.~$N$ for $\beta=1$ (Gabriel graphs). The inset in (b) is an enlargement of the 
main panel in the interval $1\leq\beta \leq2$.}
\label{Fig10}
\end{figure*}

The term ``localization transition'' we used in the previous subsection to describe the sharp decrease of $\mu$ 
at $\beta=1$ implies that we expect the eigenvectors of the adjacency matrices of BSGs to be mostly localized 
for $\beta>1$. In the following we verify this statement.

To measure quantitatively the spreading of eigenvectors in a given basis, i.e., their localization properties, 
two quantities are mostly used: (i) the information or Shannon entropy $S$ and (ii) the inverse participation 
number $I_2$. Indeed, both have been widely used to characterize the eigenvectors of the adjacency matrices 
of random network models. For the eigenvector $\Psi^m$, associated with the eigenvalue $\lambda^m$, they are given as 
\begin{equation}
S^m = -\sum_{n = 1}^N \mid \Psi_n^m \mid^2 \ln \mid \Psi_n^m \mid^2 ,
\label{S}
\end{equation}
and
\begin{equation}
I_2^m = \sum_{n=1}^N | \Psi_n^m |^4.
\label{I2}
\end{equation}
These measures provide the number of main components of the eigenvector $\Psi^m$. 
Moreover, $S^m$ allows to compute the so called entropic eigenvector localization length~\cite{I90} 
\begin{equation}
\ell = N \exp[- (S_{\tbox{GOE}} - \langle S^{m} \rangle )] ,
\label{ell}
\end{equation}
where $S_{\tbox{GOE}}$ is the average entropy of a random eigenvector with Gaussian distributed amplitudes (i.e., 
an eigenvector of the GOE) which is given by~\cite{BMHJK}
\begin{equation}
S_{\tbox{GOE}} = \psi\left(\frac{N}{2} + 1\right) - \psi\left(\frac{3}{2} \right) .
\label{SGOE}
\end{equation}
Above,  $\langle \cdot \rangle$ denotes average and $\psi(\cdot)$ is the Digamma function;
$S_{\tbox{GOE}}\approx\ln(N/2.07)$ for large $N$.

We average over all eigenvectors of an ensemble of adjacency matrices of size $N$ to compute 
$\langle S^m \rangle$, such that for each combination $(N,\beta)$ we use $5\times 10^5$ eigenvectors. 
With definition (\ref{ell}), when $\beta\to\infty$, since the eigenvectors of the adjacency
matrices of BSGs have only one main component with magnitude close to one,
$\langle S^m \rangle\approx 0$ and $\ell\approx N\exp[-S_{\tbox{GOE}}]\approx \mbox{const.}\approx 2.07$. 
On the other hand, for $\beta\to 0$, $\langle S^m \rangle\approx S_{\tbox{GOE}}$ and the fully chaotic 
eigenvectors extend over the $N$ available vertices of the BSG, so $\ell\approx N$.

Therefore, in Fig.~\ref{Fig10} we plot $\ell/N$ as a function of $\beta$ for weighted random BSGs
of sizes ranging from $N=50$ to 800. We consider both lune-based (Fig.~\ref{Fig10}(a)) and circle-based 
(Fig.~\ref{Fig10}(b)) proximity rules. As well as for the Brody parameter vs.~$\beta$ (see Fig.~\ref{Fig9}) 
here we clearly observe a sharp transition from delocalized to localized eigenvectors at $\beta=1$. 
Additionally, in the inset of Fig.~\ref{Fig10}(a) we report $\ell/N$ vs.~$N$ for $\beta=1$. There we can clearly 
see the GOE ($\ell/N\sim 1$) to Poisson ($\ell/N\sim 0$) transition in the eigenvector properties of Gabriel graphs, 
also reported through spectral properties, by the use of the $P(s)$, see the inset of Fig.~\ref{Fig9}(b).

\begin{figure}
\includegraphics[scale = .5]{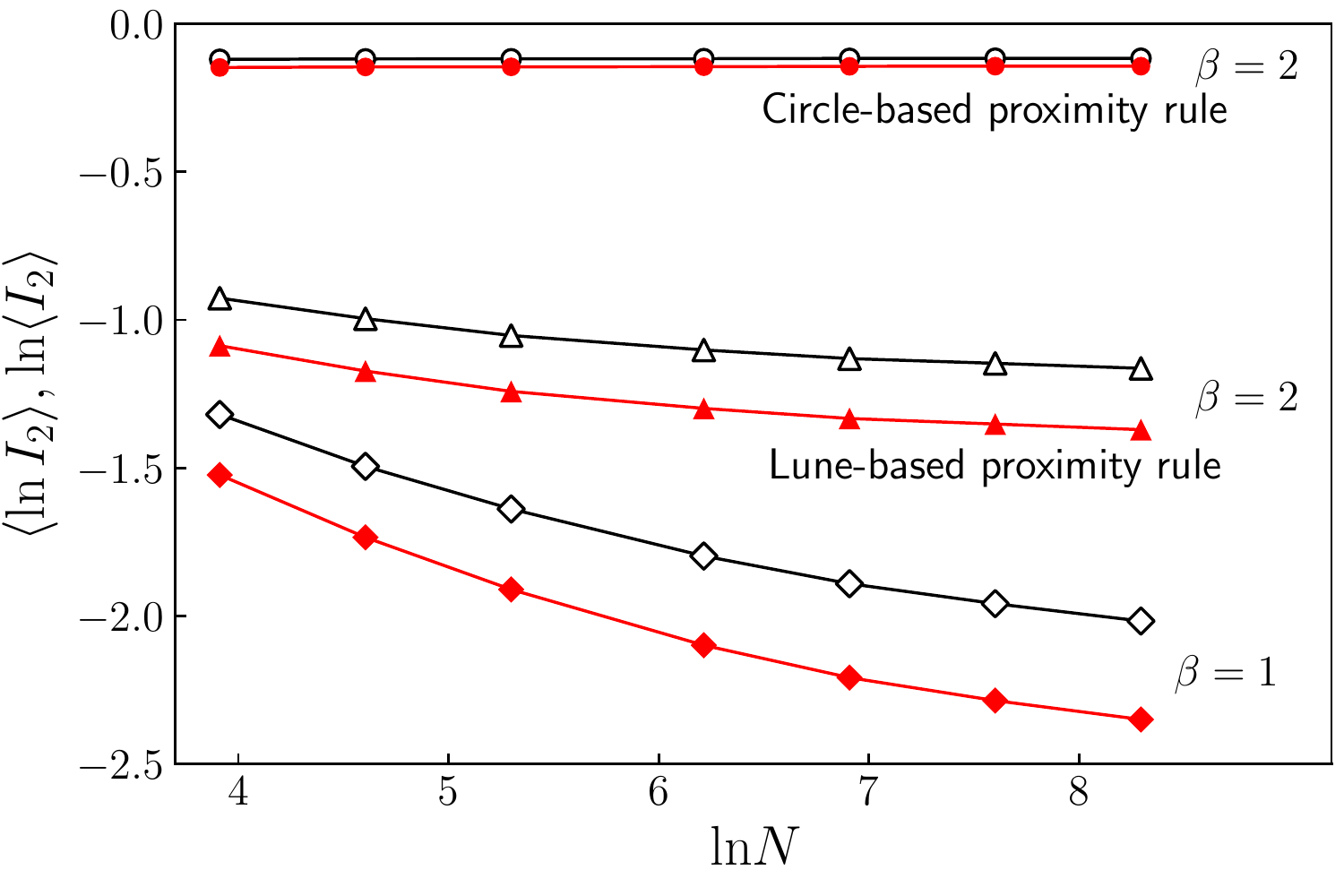}
\caption{$\langle \ln I_2 \rangle$ (red full symbols) and $\ln \langle I_2 \rangle$ (black empty symbols) 
as a function of $\ln N$ 
for Gabriel graphs (diamonds) and relative neighborhood graphs (triangles and circles for
lune-based proximity rule and circle-based proximity rule, respectively).}
\label{Fig11}
\end{figure}

Finally, we would like to add that the inverse participation number of the eigenvectors of BSGs 
shows an equivalent panorama to that reported in Fig.~\ref{Fig10} for $\ell$, so we do not show it here.
Instead, in Fig.~\ref{Fig11} we plot $\langle \ln I_2 \rangle$ and $\ln \langle I_2 \rangle$ as a function of $\ln N$ 
for Gabriel graphs ($\beta = 1$) and relative neighborhood graphs ($\beta = 2$).
The non-linear trend of the curves corresponding to $\beta=1$ and $\beta=2$ in the lune-based proximity rule
rejects the possible existence of a localization transition of the Anderson type where the eigenvectors are 
multifractal objects characterized by a set of dimensions $D_q$, where the correlation dimension $D_2$
can be extracted from the scalings $I_2^{\tbox{typ}} \propto N^{-D_2}$ or
$\langle I_2 \rangle \propto N^{-D_2}$ ($I_2^{\tbox{typ}} \equiv \exp \langle \ln I_2 \rangle$ is known 
as the typical value of $I_2$). See for example Refs.~\cite{TM16,GGG17,VMR19} where multifractality
of eigenvectors has been reported in random graph models.
Moreover the independence of both $\langle \ln I_2 \rangle$ and $\ln \langle I_2 \rangle$ on $N$
for $\beta=2$ in the circle-based proximity rule confirms that the corresponding
eigenvectors are in the localized regime; that is, $D_2\approx 0$.

\section{Conclusions}

In this paper we perform a thorough study of a particular type of proximity graphs known as 
$\beta$-skeleton graphs (BSGs). In a BSG two vertices are connected if a proximity rule, that 
depends of the parameter $\beta\in(0,\infty)$, is satisfied.
We explore the two known versions of them: lune-based and circle-based BSGs.

Our main result is the identification of a delocalization-to-localization transition
at $\beta=1$ for the eigenvectors of the adjacency matrices of BSGs for increasing $\beta$. 
It is important to stress that the localized phase corresponds to mostly isolated vertices while
the delocalized phase identifies mostly complete graphs, see e.g.~\cite{AMGM18,MAM15,VMR19}. 
We characterize the delocalization-to-localization transition by means of topological and
spectral properties; we use the standard average degree as topological measure and, within a 
random matrix theory approach, the nearest-neighbor energy-level spacing distribution 
and the entropic eigenvector localization length as spectral measures.

\begin{acknowledgments}
This work was partially supported by 
VIEP-BUAP (Grant No.~100405811-VIEP2019) and
Fondo Institucional PIFCA (Grant No.~BUAP-CA-169). 
\end{acknowledgments}


\begin{thebibliography}{99}

\bibitem{B11} 
M. Barth\'el\'emy, 
Spatial networks, 
Phys. Rep. {\bf 499}, 1 (2011).

\bibitem{B18}
M. Barthelemy, 
{\it Morphogenesis of Spatial Networks},
Lecture Notes in Morphogenesis
(Springer International Publish, 2018).

\bibitem{DGKJDR}
D. G. Kirkpatrick and J. D. Radke, 
{\it A framework for computational morphology}
(IBM Thomas J. Watson Research Division, 1984).

\bibitem{GTT}
G. T. Toussaint,
The relative neighborhood graph of a finite planar set,
Pattern Recognition {\bf 12}, 261 (1980).

\bibitem{BOD12}
D. S. Bassett, E. T. Owens, K. E. Daniels, and M. A. Porter,
Influence of network topology on sound propagation in granular materials,
Phys. Rev. E {\bf 86}, 041306 (2012).

\bibitem{OH14}
T. Osaragi and Y. Hiraga,
Street network created by proximity graphs: Its topological structure and travel efficiency.
In: Proceedings of the 17th Conference of the Association of Geographic Information Laboratories for Europe 
on Geographic Information Science (AGILE2014), 
Huerta, Schade, Granell (Eds), pp. 1-6, Castell\'on, Spain (2014).

\bibitem{estrada}
E. Estrada and M. Sheerin,
Random neighborhood graphs as models of fracture networks on rocks: Structural and dynamical analysis,
Appl. Math. Comp. {\bf 314}, 360 (2017).

\bibitem{KRRRS}
K. R. Gabriel and R. R. Sokal,
A new statistical approach to geographic variation analysis,
Systematic Zoology {\bf 18}, 259 (1969).

\bibitem{metha}
M. L. Mehta,
{\it Random matrices} (Elsevier, Amsterdam, 2004).

\bibitem{ES15}
E. Estrada and M. Sheerin, 
Random rectangular graphs,
Phys. Rev. E {\bf 91}, 042805 (2015).

\bibitem{AMGM18}
L. Alonso, J. A. Mendez-Bermudez, A. Gonzalez-Melendrez, and Y. Moreno,
Weighted random-geometric and random-rectangular graphs: Spectral and eigenvector properties of the adjacency matrix,
J. Complex Networks {\bf 6}, 753 (2018).

\bibitem{B73}
T. A. Brody,
A statistical measure for the repulsion of energy levels,
Lett. Nuovo Cimento {\bf 7}, 482 (1973).

\bibitem{B81}
T. A. Brody, J. Flores, J. B. French, P. A. Mello, A. Pandey, and S. S. M. Wong,
Random-matrix physics: spectrum and strength fluctuations,
Rev. Mod. Phys. {\bf 53}, 385 (1981).

\bibitem{BJ07}
J. N. Bandyopadhyay and S. Jalan,
Universality in complex networks: Random matrix analysis,
Phys. Rev. E {\bf 76}, 026109 (2007).

\bibitem{JB08}
S. Jalan and J. N. Bandyopadhyay,
Random matrix analysis of network Laplacians,
Physica A {\bf 387}, 667 (2008).

\bibitem{ZYYL08}
G. Zhu, H. Yang, C. Yin, and B. Li,
Localizations on complex networks,
Phys. Rev. E {\bf 77}, 066113 (2008).

\bibitem{J09}
S. Jalan,
Spectral analysis of deformed random networks
Phys. Rev. E {\bf 80}, 046101 (2009).

\bibitem{JB07}
S. Jalan and J. N. Bandyopadhyay,
Random matrix analysis of complex networks,
Phys. Rev. E {\bf 76}, 046107 (2007).

\bibitem{MAM15}
J. A. Mendez-Bermudez, A. Alcazar-Lopez, A. J. Martinez-Mendoza, F. A. Rodrigues, and T. K. DM. Peron,
Universality in the spectral and eigenvector properties of random networks,
Phys. Rev. E {\bf 91}, 032122 (2015).

\bibitem{DGK16}
C. P. Dettmann, O. Georgiou, and G. Knight,
Spectral statistics of random geometric graphs,
Europhys. Lett. {\bf 118}, 18003 (2017).

\bibitem{I90}
F. M. Izrailev,
Simple models of quantum chaos: Spectrum and eigenvectors,
Phys. Rep. {\bf 196}, 299 (1990).

\bibitem{BMHJK}
B. Mirbach, H.J. Korsh, 
Annal. Phys. {\bf 265}, 80 (1998).

\bibitem{TM16}
K. S. Tikhonov and A. D. Mirlin,
Fractality of wave functions on a Cayley tree: Difference between tree and locally treelike graph without boundary,
Phys. Rev. B {\bf 94}, 184203 (2016).

\bibitem{GGG17}
I. Garc\'ia-Mata, O. Giraud, B. Georgeot, J. Martin, R. Dubertrand, and G. Lemarie,
Scaling theory of the Anderson transition in random graphs: Ergodicity and universality,
Phys. Rev. Lett. {\bf 118}, 166801 (2017).

\bibitem{VMR19}
D. A. Vega-Oliveros, J. A. Mendez-Bermudez, and F. A. Rodrigues,
Multifractality in random networks with power--law decaying bond strengths,
Phys. Rev. E {\bf 99}, 042303 (2019).


\end{thebibliography}
\end{document}